# Panoramic SETI: overall mechanical system design

Aaron M. Brown[*,a]; Michael L Aronson[g]; Shelley A. Wright[a,b]; Jérôme Maire[a]; Maren Cosens[a,b]; James H. Wiley[a,b]; Franklin Antonio[f] ; Paul Horowitz[e]; Rick Raffanti[c]; Dan Werthimer[c,d]; Wei Liu[c,h]
[a]Center for Astrophysics & Space Sciences, University of California San Diego, USA; [b]Department of Physics, University of California San Diego, USA; [c]Department of Astronomy, University of California Berkeley, CA, USA; [d]Space Sciences Laboratory, University of California Berkeley, CA, USA; [e]Department of Physics, 17 Oxford St, Cambridge, MA, 02138, USA; [f]Qualcomm Research Center, 5775 Morehouse Dr, San Diego, CA, USA; [g]Electronic Packaging Man, Encinitas, CA, USA; [h]Institute of RF- & OE-ICs, Southeast University, China

## ABSTRACT

PANOSETI (Pulsed All-Sky Near-infrared Optical Search for Extra Terrestrial Intelligence) is a dedicated SETI (Search for Extraterrestrial Intelligence) observatory that is being designed to observe 4,441 sq. deg. to search for nano- to milli-second transient events. The experiment will have a dual observatory system that has a total of 90 identical optical 0.48 m telescopes that each have a 99 square degree field of view. The two observatory sites will be separated by 1 km distance to help eliminate false positives and register a definitive signal. We discuss the overall mechanical design of the telescope modules which includes a Fresnel lens housing, a shutter, three baffles, an 32x32 array of Hamamatsu Multi-Photon Pixel Counting (MPPC) detectors that reside on a linear stage for focusing. Each telescope module will be housed in a triangle of a 3$^{rd}$ tessellation frequency geodesic dome that has the ability to have directional adjustment to correct for manufacturing tolerances and astrometric alignment to the second observatory site. Each observatory will have an enclosure to protect the experiment, and an observatory room for operations and electronics. We will review the overall design of the geodesic domes and mechanical telescope attachments, as well as the overall cabling and observatory infrastructure layout.

## 1. INTRODUCTION

The Pulsed All-Sky Near-infrared Optical Search for Extra Terrestrial Intelligence (PANOSETI) observatory is designed to house multiple 0.48 m Fresnel telescopes (i.e., modules) tiled in a spherical geometry to observe 4,441 sq. degrees instantaneously at optical wavelengths between 300 – 850nm[1,2]. Each module will be arranged in a geodesic dome structure that is housed in a clam shell outer enclosure. Each telescope module will consist of a Fresnel lens housing, an outer housing that will include baffling, and a removable electronics package. PANOSETI has two identical observatory systems that are located at two different sites separated by 1 km to eliminate false positives from Cherenkov showers and any other earth-bound phenomenon[6]. PANOSETI is designed to explore a unique time-domain in astrophysics ranging from nanosecond to second resolution to search for astrophysics transients and potential technosignatures[3].

PANOSETI has several unique optical and mechanical design considerations: a mass-production run of ninety 0.48 m Fresnel lens telescopes, electronics packaging of 1GHz MPPC optical arrays, two 6 m diameter geodesic dome, and two automatic enclosure with safety considerations. In this paper we discuss and review the latest design of the opto-mechanical components of our prototype units and the production design of this system.

---

[*] ambrown@ucsd.edu; phone 1 858 534-5601; oirlab.ucsd.edu

## 2. PROTOTYPE TELESCOPES

The final PANOSETI observatory concept is planned for development at Palomar observatory. For prototyping and project development, we deployed two remote operated prototype modules at Lick Observatory. The prototypes are currently housed in the Astrograph dome, which has been updated to meet the needs of the two apertures (Figure 1). The prototype modules each contain one 0.48 m Fresnel lens assembly that is supported by four aluminum 6061-T6 square rods that are 37.9 mm by 37.9 mm and 670.6 mm long. Each rod has had material removed for lightening purposes and is black anodized to minimize light reflections. Rods are connected to a support platform that is 6061-T6 black anodized aluminum with a diameter of 503 mm and a thickness of 12 mm. Attached to one side of the support platform is the stepper motor bracket and stepper motor. The remaining focus stage and electronics components for the detector plane (Figure 2 a) are on the opposite side of the support platform. The electronic packages include a mother board that drives four custom readout boards or quadrant boards (quabo)[7], that mount the Hamamatsu Multi-Photon Pixel Counting (MPPC) detectors creating a 32x32 focal plane array. Both telescopes are housed inside of a Dobsonian box, made from high density polyethylene (HDPE) sheets in black (12.7 mm thick) with and overall size 631.8 mm tall by 631.8 mm wide and 711.2 mm deep, that allows altitude adjustment (Figure 2b). On-sky observations have been occurring for the last 7 months to allow software development. These prototype modules are a major milestone in the development of the final production modules that will be housed in their own purpose-built observatories.

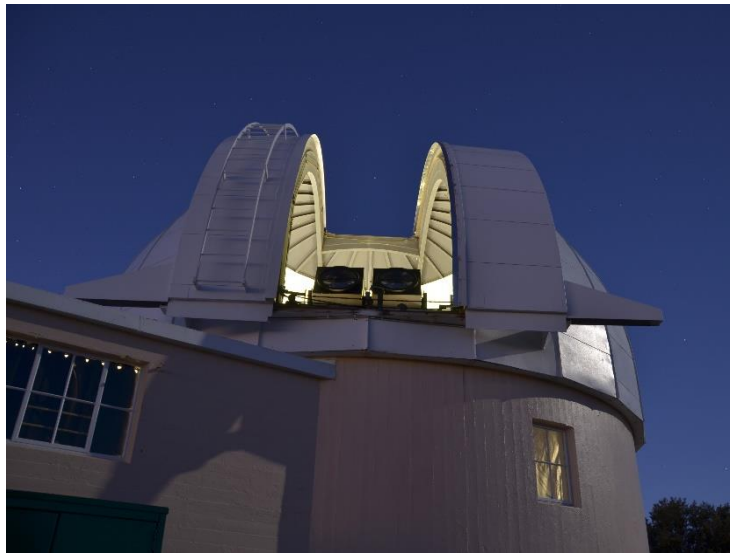

Figure 1: A photograph of the Astrograph dome at Lick Observatory in operation in March 2020 with the recently commissioned two 0.48 m PANOSETI prototype telescopes. Both PANOSETI telescopes are side-by-side and observe simultaneously the same field of view for signal confirmation (Photograph by Laurie Hatch[8]).

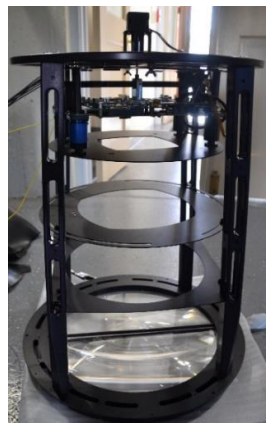 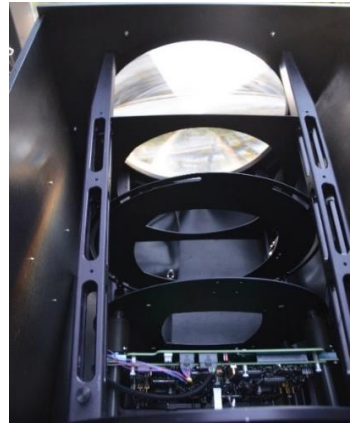

(a)          (b)

Figure 2: (a) PANOSETI prototype module. From the bottom up, the Fresnel lens assembly with an outer diameter of 606 mm is attached to four aluminum 6061-T6 square support rods that are 37.88 mm by 37.88 mm and 670.61 mm long connected to a support platform made from 6061-T6 black anodized aluminum with a diameter of 502.89 mm and is thickness of 12 mm. On the top side of the support platform a 3D printed stepper motor bracket made from Polyethylene terephthalate glycol (PETG) can be seen with the focus stepper motor attached to it. Below the support platform, four guide rods are attached, with a baffle attached to the end of the guide rods. There are four IGUS FJUM-02-16 linear guide bearings for the detector plane to translate, and the bearings are attached to the main motherboard of the electronics package. Attached to the back side of the mother board is a focus bracket with an IGUS drylin® low profile trapezoidal lead screw nut. The lead screw can be seen going from the lead screw nut to a coupling that is attached to the stepper motor. This system allows for the detector plane to translate in order to focus. Two other baffles are attached to the support rods using 90-degree brackets. (b) The PANOSETI prototype module installed in the Dobsonian box made from high density polyethylene (HDPE) sheets in black (12.7 mm thick) with an overall size 631.8 mm tall by 631.8 mm wide and 711.2 mm deep, that allows altitude adjustment. From the top of the picture the Fresnel lens assembly can be seen attached to the 4 support rods and the support platform at the other end. On the bottom the electronics package can be seen attached to the guide rods on each side leading to the lower baffle. This internal support structure is wrapped in neoprene to minimize light scatter and then a lid is placed on top fully enclosing the system.

## 3. TELESCOPE MODULES

The prototype telescopes were essential for learning about the lens assembly and electronics packaging. The final observatory requires a compact telescope housing and ease of accessibility to the electronics for installation and servicing. We describe the final design of our production telescope modules that are designed for installation in a geodesic dome (Section 4).

The production telescope modules are defined as an aperture unit that contains a 0.48 m Fresnel lens (f/1.3), a detector, and electronics focal plane[1]. These modules include the opto-mechanical assembly for the Fresnel lens, main telescope housing, and the electronics packaging, which include the detector focal plane arrays. Figure 3 shows the module rendering and a breakdown of individual components.

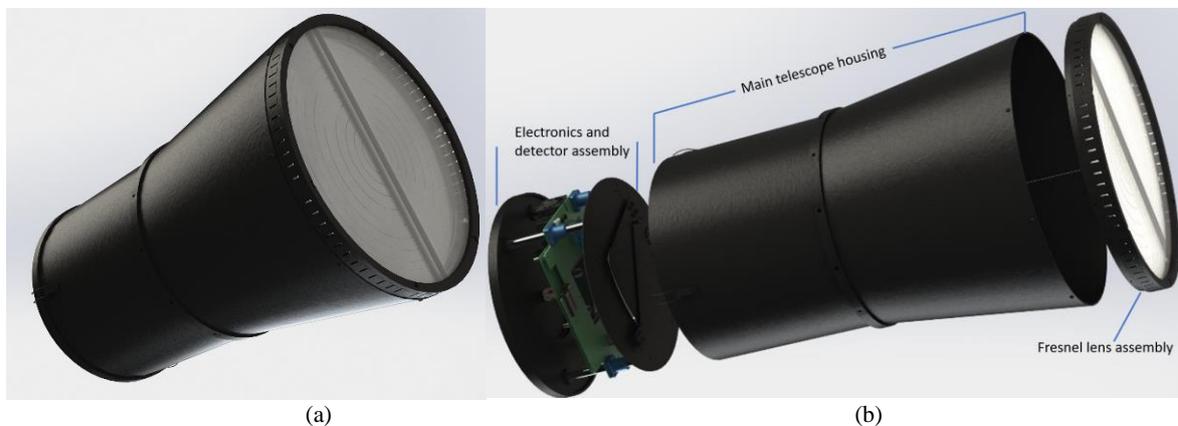

(a) (b)

Figure 3: (a) The assembled telescope module. The length of the module is 798 mm. The bottom cylinder is 410 mm and top angled-cylinder is 495 mm. The effective lens diameter when assembled is 460 mm. (b) Exploded view of the three major components can be seen from left to right: the electronics and detector assembly, main telescope housing, and the Fresnel lens assembly. All electronics and moving components are easily removed for servicing and installation.

### 3.1 Fresnel Lens Assembly

The Fresnel lens are 0.48 m in diameter and are enclosed in a 6061-T6 aluminum frame that is designed to be as small as possible with an outer diameter of 495 mm while protecting and supporting the rather flexible lens which is 1.8 mm thick. The lens is also protected with an outer shield of plexiglass to prevent any contact. The plexiglass shield is 2.39 mm thick and has a diameter of 490 mm. The outer frame is 495 mm and has slots around the diameter to provide air flow to eliminate moisture build up on the lens and protective cover. This slot design for air flow has been tested on the prototype modules in the field and has been found to be prevent condensation. The lens is held in place by eight M4 shoulder screws evenly spaced around its perimeter. To support the lens, two aluminum 6061-T6 support bars are used, one below and one above, that run across the center of the lens. The support bars are rectangular (3.2 mm by 9.5 mm) to minimize aperture blockage. Above the support bars is the acrylic lens to help isolate the Fresnel lens from direct exposure to the elements. To encapsulate these components, a top aluminum 6061-T6 retainer ring is used, and is screwed into the main lens frame with six M4 panhead screws. All aluminum components are bead blasted and black anodized to obtain a matte black finish. Figure 4 shows how each of these lens assembly components are combined in assembly.

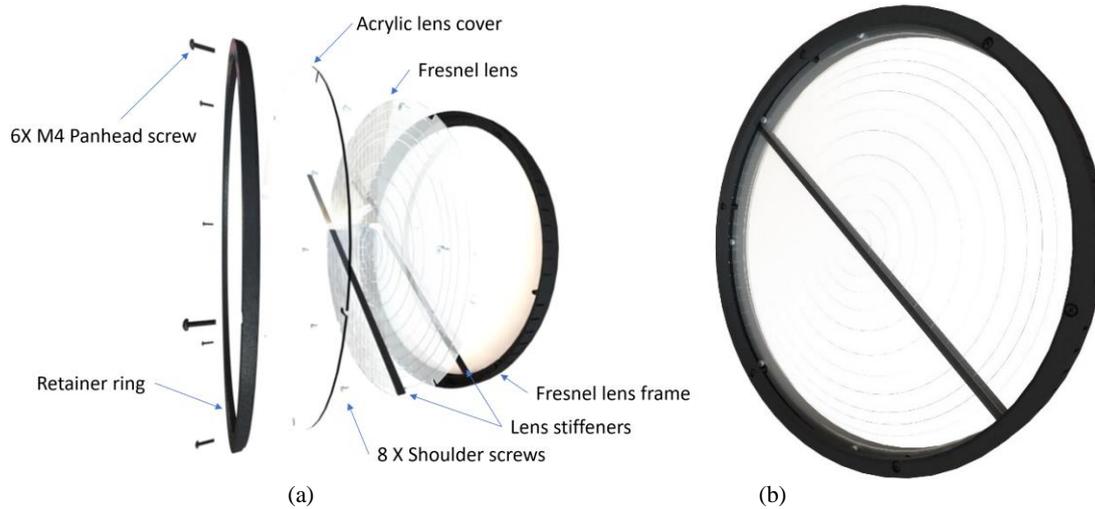

(a)　　　　　　　　　　　　　　　　　　　　(b)

Figure 4: (a) Exploded view of the Fresnel lens housing. From left to right: six black-oxide 18-8 stainless steel pan head Philips screws secure the retainer ring to the Fresnel lens frame. Next, the retainer ring made from 6061-T6 aluminum with an inner diameter of 460 m, an outer diameter of 495 mm and a height 10.5 mm encapsulates the components to the Fresnel lens frame. Below the retaining ring is the acrylic protective cover that is 2.39 mm thick with an outer diameter of 490 mm. The two 6061-T6 aluminum support bars sandwich the Orafol Fresnel lens. The top support rod is 9.53 mm tall by 3.18 mm wide and 490.2 mm long. The bottom support rod has the same height and width but a length of 481 mm. Between the support rods is an Orafol SC 214 positive Fresnel lens with a focal length of 607.8 mm, a thickness of 1.8 mm and a diameter of 480 mm. All aluminum components are bead blasted and black anodized to obtain a matte black finish. (b) The Fresnel lens assembled and viewed from the top with a height of 43 mm and a maximum diameter 495 mm.

## 3.2 Telescope Module Housing

The electronics package and lens assembly are mounted to the main housing at opposite ends. The main housing is comprised of a lower housing with an inner diameter of 392 mm and a height of 324 mm, a mounting flange, and an upper conical housing with a top inner diameter of 490 mm and a lower inner diameter of 396 mm. The overall height of the conical housing is 370 mm (see Figure 5 for component view). The upper conical housing and the lower housing are made from sheets of 5052-H32 aluminum 3.17 mm and 2.38 mm thick respectively. Each of these pieces are rolled into shape and seam welded. The mounting flange is also made form 5052-H32 aluminum with and outer diameter of 412 mm and a thickness of 25.4 mm. The flange has twelve M8 mounting holes spaced equally around the outer diameter. The middle baffle is integrated into the mounting flange with an inner diameter of 275 mm. The lens assembly is mounted using eight M4 button head hex drive black-oxide alloy steel screws to the top upper conical housing. The electronics package is secured to the lower housing using spring loaded clips. This allows not only for easy extraction of the electronics module, but allows for the system to be rotated for alignment purposes.

　　　　This entire structure is constructed using aluminum to keep the system as light as possible. All of the body components are welded together to increase stiffness of the structure and minimize mounting points. The conical housing is angled at 82.6° to allow for a larger lens, while decreasing the diameter from 492 mm at the top to 396 mm at the bottom end. Reducing the overall footprint of the individual telescopes minimizes the geodesic dome array, allowing the use of off-the-shelf dome structures. There are two circular baffles integrated into the housing to mitigate scattered light off the Fresnel lens. One is welded into the conical housing and the other is machined into the mounting flange as previously stated. In the prototype design, the baffles were designed to be a combination of a round lens and the square detector to maximize baffling effect. This shape is called a superellipse, but was not able to be maintained in the final design for the upper two baffles, because the electronics module is rotatable. However, the final baffle before the detector does maintain the superellipse shape to baffle as much light as possible. The upper baffle not only helps form the conical housing, but also increases the rigidity of the unit. The mounting ring has the second baffle machined into its center, also adding rigidity to the housing. A cross section of the complete module can be seen in Figure 6. The mounting flange has twelve evenly spaced mounting M8 holes around its diameter to make the complete system modular for small scale testing to final production.

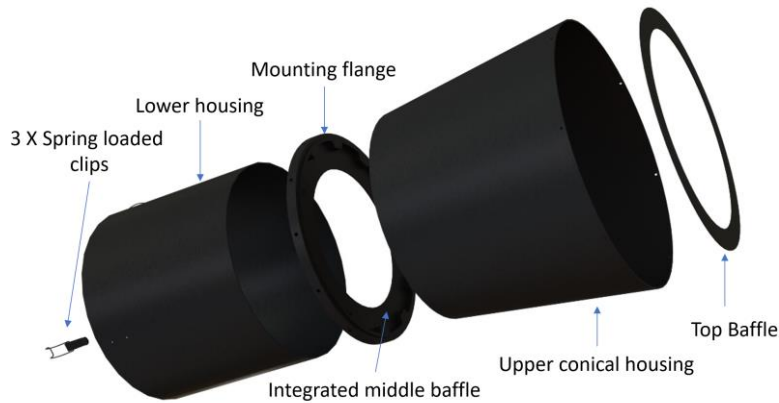

Figure 5: Exploded view of the module housing. Components from right to left: top baffle ring with an inner diameter of 364 mm, an outer diameter of 441 mm and a thickness of 2 mm is tack welded to the upper housing 196 mm from the top of the conical housing. The conical housing has a top inner diameter of 492 mm and a bottom inner diameter of 396 mm with a total height of 370 mm. The conical housing has an angle of 82.6° and is 3.17 mm thick and is welded to the mounting flange. This is followed by the mounting flange which incorporates the middle baffle with an inner diameter of 275 mm and an outer diameter of 412 mm. The mounting flange has 12 evenly spaced M8 holes. The lower housing, which is welded to the mounting flange, is 2.38 mm thick with an inner diameter of 392 mm. All of these components are 5052-H32 aluminum for ease of welding and cold formability. The outside of the assembly will be matte black powder coated and the inside will be a polyurethane paint design for optics. The final component in the assembly is the nickel-plated iron spring-loaded draw-toggle latch that is used to hold the electronics and detector assembly inside the housing. The mounting flange will be the connection point to the geodesic dome structure.

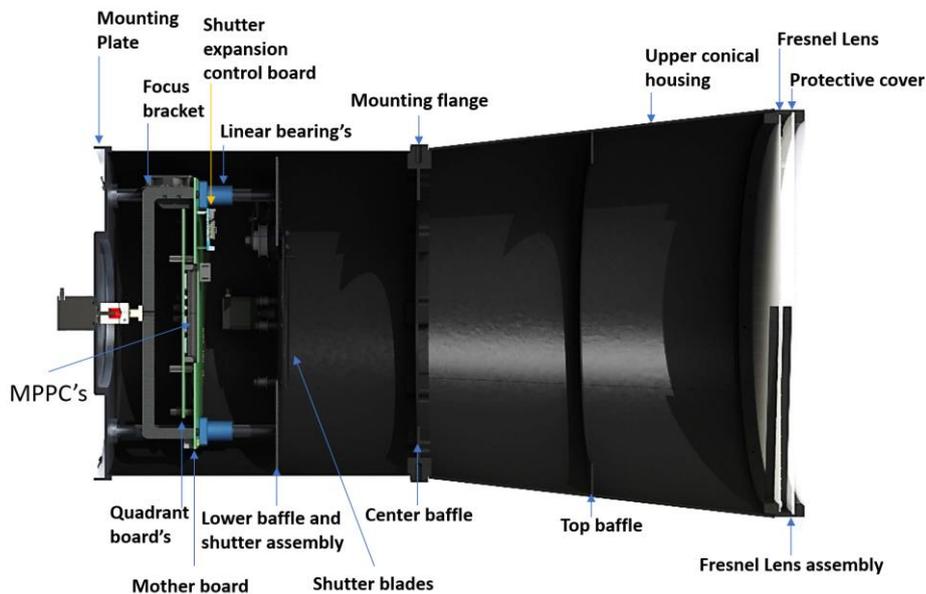

Figure 6: A cross section of the complete module including the Fresnel lens assembly at the right, the three baffles seen as labeled, the main housing consisting of the upper conical housing, top baffle, mounting flange with incorporated center baffle, and lower housing; and to the left the removable electronics package with the focus assembly, mother board, quadrant boards, and shutter assembly.

## 3.3 Electronics package

The electronics package is the final major component in the telescope assembly seen in Figure 7(a). It can be broken down into three sub-assemblies: the shutter, the detector and electronics, and the focus stage. Starting from the top, the shutter assembly is designed as a final safety measure to protect the detectors from potentially damaging sunlight. The shutter system is controlled by an expansion card on the motherboard that has an integrated logic chip. The shutter has two opposing blades machined from 5052-H32 aluminum. Each shutter blade is bolted to a coupler with four M5 pan head bolts. The coupler is made from 45 carbon steel. Below the coupler is a pillow block bearing followed by a 60-tooth GT2 synchronous aluminum timing pulley. The pulleys are connected with a standard GT2 timing belt, making them gravity neutral to minimize the force needed to open and close them. One shutter blade stack has a limit switch bracket, that depresses two limit switches as the blades open and close. Also attached to the switch bracket is a spring that will pull the shutters closed if light is detected or if power is lost. One switch is for the open position and one for the closed position. There are four idler bearings which maneuver the GT2 timing belt around the baffle opening so that the light path is not interrupted. The belt also connects to a 16-tooth synchronous timing pulley that is attached to a NEMA 17 stepper motor used to open and close the shutter blades in a controlled manner. With a gear ratio of 3.75 a smaller stepper motor can overcome the friction and closing-spring forces needed. If light is detected or power is lost the spring in the linkage will quickly close the blades and block out any focused light to the detectors. A top view of the shutter system closed can be seen in figure 7 (b). All of these components that are mounted to the lower baffle are machined from 5052-H32 aluminum. The two shutter blades, lower baffle, blade couplers, and switch bracket are all matte black powder coated to decrease light reflection and provide a curable coating.

The shutter assembly is attached to four 6061-T6 aluminum guide rods down to the mounting plate. On the back of the mounting plate is a second NEMA 17 stepper motor. The mounting plate is machined from 6061-T6 and will be bead blasted and black anodized. On the top side of the mounting plate the motor shaft connects to a coupler that attaches to an M5 threaded rod on its opposing side. The threaded rod threads into an IGUS Dylin® low profile trapezoidal lead screw nut (p/n JFRM-0913M5) that attaches to the focus bracket. The focus bracket is also made from 6061-T6 aluminum and will be bead blasted and black anodized for a matte black finish. The focus bracket has a cooling fan bracket with a fan attached. The focus bracket bolts to the motherboard with four M5 bolts, two on each side.

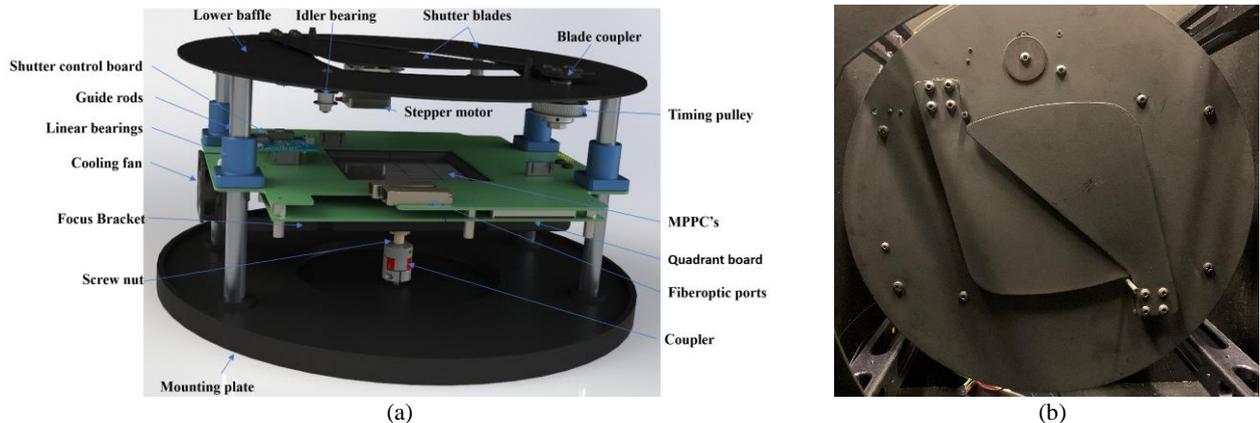

(a) (b)

Figure 7(a) Electronics and detector module of the aperture assembly. On top, the shutter assembly can be seen. It includes two opposing shutter blades. Each shutter blade is bolted to a coupler with four M5 pan head bolts. A bolt is used to connect a pillow block bearing followed by an aluminum 60-tooth timing pulley. There are four idler bearings which maneuver the belt around the baffle opening to keep it out of the light path. The belt then links to a NEMA 17 stepper motor with a 16-tooth timing gear. All of these components are mounted to the lower baffle. The two shutter blades, lower baffle, blade couplers, and switch bracket are all matte black powder coated to decrease light reflection and provide a curable coating. The shutter assembly is attached to four guide rods down to the mounting plate. On the back of the mounting plate (not seen here) is a second NEMA 17 stepper motor that is mounted directly to the mounting plate. On the top side of the mounting plate the shaft for the motor connects to a coupler that attaches to an M5 threaded rod on its opposing side. The threaded rod threads into an IGUS Dylin® screw nut that attaches to the focus bracket. The focus bracket also has a cooling fan bracket and fan attached to the left. The focus bracket bolts to the motherboard. The detector and electronics assemblies can be seen here but will be further described in Figure 8 with the exception of the shutter control board which is shown here attached to the top side of the motherboard. (b) The shutter assembly top view can be seen here showing profile of both blades closed blocking all focused light from the electronics assemblies.

The next major system is the detector and electronics assembly seen in Figure 8. The largest component in this assembly is the motherboard which is the main interface to external systems. It has no logic processors but mechanically it serves as the backbone structurally and for software it provides communication between the entire system. It has three 8-pin Molex expansion card connections for any future additions. One of these slots is already being used by the controller card for the shutter, leaving two for future expansion. There are two fiberoptic ports, one for data and one using White Rabbit[5] precision timing protocol. Several other Molex connectors are used to control the focus stage motor, a cooling fan, and an optical limit switch. A 6-pin Samtec power connector is used to distribute power throughout the system. Top view seen in Figure 8 (a). Connected to the mother board from below are four custom readout boards that make use of 64-bit Application-Specific Integrated Circuit (ASIC) Weeroc Maroc-3A's that are capable of pulse shaping and providing trigger detection of individual pulses[3] seen in Figure 8(b). Each of the boards has four 8x8 arrays of Hamamatsu silicon photomultipliers (SiPM) detectors (p/n S13361-3050AE-08) that will be tiled on the optical axis to create a continuous 32x32 pixel field of view[3]. For further description of the electronics and internal software design see Wei et al., this conference.

There are four IGUS FJUM-02-16 pillow block bearings attached to the motherboard and guided on precision 16 mm IGUS rods attached to the base mounting plate of the electronics module. There is also a focus bracket attached to the motherboard that serves multiple purposes including encapsulating a leadscrew nut, a mounting point for a fan bracket and fan, and finally to add stiffness to the assembly. The leadscrew nut is an IGUS Dylin® low profile trapezoidal lead screw nut (p/n JFRM-0913M5). An M5 threaded rod is connected to the lead screw nut and a dampening coupler on the other end minimizes shock through the system during focus calibrations. The coupler is attached to a stepper motor (p/n 17HM13-0316S) which is secured to the base mounting plate. The stepper motor has a step angle of 0.9° and the guide rod has a thread pitch of 0.8 mm allowing for a focal step of 2 microns. The electronics module was designed to be easily removable from the observatory housing for maintenance and cleaning. To quickly secure the modules, spring loaded clips are used to secure the module to the housing as seen in Figure 5. The entire electronics package is designed to rotate to any field position angle on-sky and be secured. All data and power are contained to the electronics module and will be connected to passthrough connectors on the base mounting plate. A more in-depth look at the focal plane electronics, timing, and network protocols can be reviewed in "Wei et al, this conference"[7]

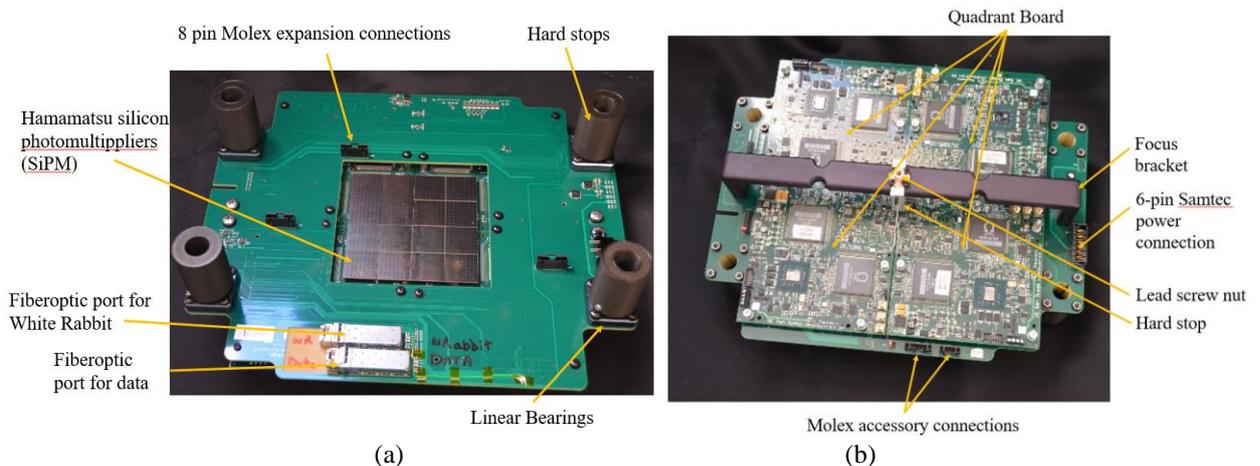

(a) (b)

Figure 8: Electronics and detector module of the aperture assembly. (a) Top of the motherboard with the sixteen Hamamatsu silicon photomultipliers (SiPM) 8x8 detector arrays seen at the center. This array is distributed between four quadrant boards, bottom seen in Figure 8 (b), with a smaller 2x2 arrays on each. On the motherboard, three 8-pin molex expansion card slots, a fiberoptic port for White Rabbit[5], and a fiberoptic data port are seen. There are four linear bearings attached to the sides of the motherboard with a 3d printed PETG hard stop attached to each. This hard stop limits against the back end of the lower baffle. (b) Bottom of the electronics assembly showing the focus bracket across the motherboard with the lead screw attached at its middle. A 6061-T6 aluminum hard stop can be seen attached to the focus bar to protect the quadrant boards from the lead screw. Four custom readout boards, or quadrant boards, can be seen attached to the underside of the motherboard. The 8x8 detector array that connects to the opposite side of the quadrant boards can be seen in Figure 8(a). There is a 6-pin Samtec power connector that provides power to the entire module and several Molex accessory connections that control the cooling fan and optical stop.

# 4. OBSERVATORY DESIGN

With tens of telescope modules being used per observatory, an innovative method needed to be devised to support each module and point each aperture to minimize on-sky field overlap. Using a geodesic dome and dividing it into triangular segments allows us to mount each module while allowing an efficient on-sky field coverage[4]. This method also has the benefit of being the smallest possible footprint, which in turn helps with selecting the protective observatory enclosure. A smaller footprint also reduces cost in construction at the remote and high-altitude observatory sites. The PANOSETI experiment has the added challenge of requiring to observe the entire sky while in operation. It is common for traditional observatories to use domes as a protective structure, but most of these designs have a single slit that only allows a single observational direction at a time. For the PANOSETI observatory, a protective structure needs to be utilized that allows complete protection during the day and full sky visibility at night. This entire observatory structure needs to maintain a minimal square footage. To accommodate this, a clam shell dome was selected to protect the PANOSETI instrumentation.

## 4.1 Geodesic dome

Each module will be secured to a subdivided triangle of a truncated spherical icosahedron[4] structure with the ability to have their viewing direction adjusted slightly to help overcome manufacturing tolerances and to help match viewing locations to the corresponding telescope at the second facility. The modules need to be attached to a structure that is both strong to flexure and wind, and that will have the smallest footprint and allow sky orientation adjustability. With these constraints a geodesic dome is a compact and sturdy structure that meets these requirements[4]. To mount all 45 modules, a tessellation frequency 3 icosahedron geodesic dome subdivided into triangles was selected. The geodesic dome has a side view diameter of 5.011 m with a top view diameter of 4.540 m. The overall diameter of the geodesic dome with the modules mounted is 4.988 m with a height of 1.827 m. Figure 9 (a) shows the mounting configuration of 45 modules positioned in the geodesic dome. A single aperture has a field of view of 9.95° x 9.95° and with forty-five modules we are capable of simultaneously observing 4,441 sq. deg. In Figure 9 (b) a view of the on-sky mapping can be seen using the forty-five modules from one observatory.[6] The secondary observatory has a matching field of regard for signal confirmation, see Maire et al., this conference[6].

    The geodesic dome will be made with steel tubes that are pressed flat at each end. At the center of each flat a hole is drilled. The drilled holes are where other tubes will be bolted together to make hubs and define each triangle plane in the geodesic dome. With the use of a tessellation frequency 3 geodesic dome, only three different tube lengths are required, greatly minimizing the complexity and cost. The tube lengths that will be required are 0.9272 m, 1.0224 m, and 1.0725 m. The angles that will be needed will be bent into the tube ends in order to create the geodesic dome frame. The final geodesic dome will have a pentagon center and five hexagons with each hexagon utilizing one side of the pentagon, and then two other triangles to connect each hexagon completing the final skirt as shown in Figure 9 (a).

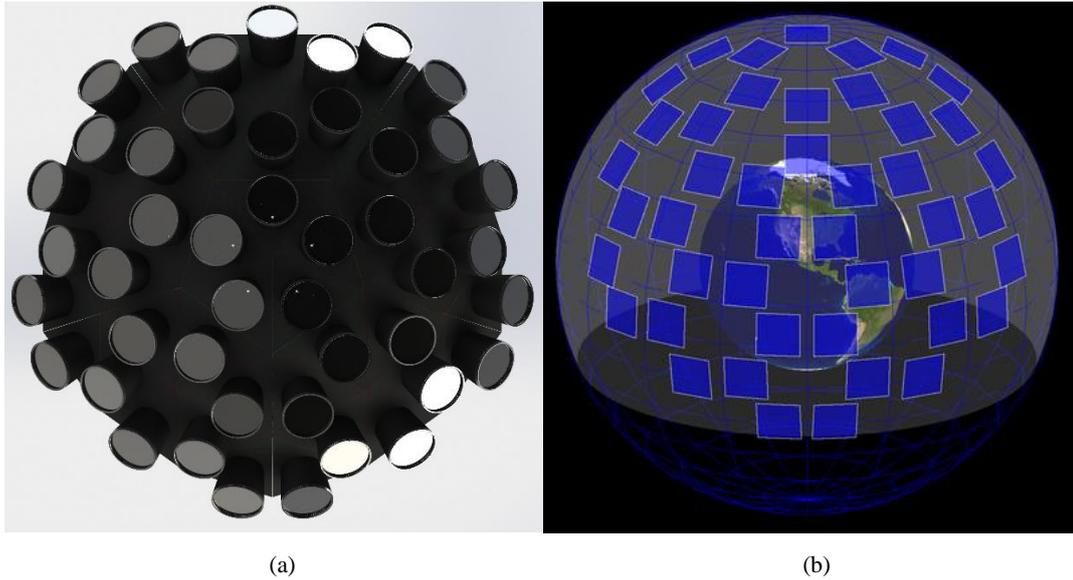

(a) (b)

Figure 9: (a) The mounting configuration incorporating 45 modules into a geodesic dome. The overall diameter of the geodesic dome with the modules is 4.988 m with and height of 1.827 m. On the back side of the modules there is an optimal clearance of 53.87 mm. (b) The on-sky projection of the observatory with forty-five modules utilizing a tessellation frequency of 3 in the geodesic dome. The secondary dome will match the field of view coverage for source confirmation. [6]

### 4.2 Observatory Building

The complete tessellation 3 geodesic dome assembly will need to be protected from the elements and enclosed during the day to keep electronics and detectors protected. To accommodate this, a clam shell dome design that opens 180° has been selected as the enclosure structure. The clam shell dome has a height of 4.33 m when closed and a diameter of 6.35 m. The dome consists of a base and five shutters that overlap to close the dome. When the dome opens, three of the shutters open one direction and the other two rotate down the opposite direction. The shutters all pivot at the same point 1.17 m above the base of the dome structure. The shutters are powered by electric motors and a pulley system that have the ability to be remotely operated. The clam shell is installed on the concrete structure that is the second story and roof of the observatory. A concept design of this off-the-shelf dome placed on top of a structure can be seen in Figure 10 (a). An interior top view of the observatory's first floor can be seen in Figure 10 (b).

The observatories are designed to be remotely operated, but will have space on the first floor for work areas, equipment, and storage. The floor height of the first floor will be 2.746 m with a floor space of 21.14 $m^2$. An enclosed stairwell will be utilized to ensure that there is no light leaked from the first floor to the instruments with a door on the lower level. The stairwell will cause a large opening in the second floor where either railing or a removable floor will be used. A double entrance door is used to prevent light leakage to other instruments at the two sites. The double entry doors are a height of 2.14 m and a width of 1.84 m to easily move equipment in-and-out of the observatory. The second floor will house the geodesic dome with installed telescope modules. The geodesic dome structure resides off the floor by 1.63 m to allow access to the modules underneath for maintenance and so the telescopes clear the structure. The geodesic dome has a support structure that attaches to the floor that makes it ridged and supports its weight. Power supplies are placed on the second floor equally spaced around the geodesic dome to minimize cabling and wiring lengths to each telescope. All fiber optic data and White rabbit signal cables are routed to the first floor for system communication and data storage in a stationary electronics rack. An internal design concept of the first floor can be seen in Figure 10 (b). The main structure has an inner diameter of 4.8 m and a height of 3.048 m. The overall structure has a height of 7.1277 m, with stairs leading to the second story.

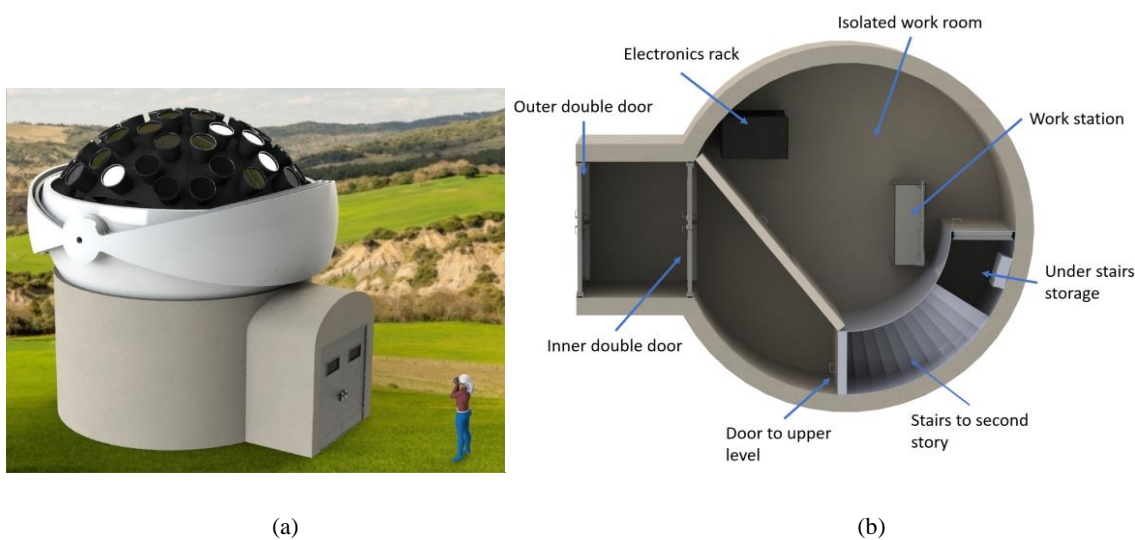

(a) (b)

Figure 10: (a) Exterior view of conceptual design for the PANOSETI observatory building which houses forty-five telescopes. The outer diameter of the concrete structure is 5.3 m with a height of 3.048 m. The overall height of the observatory is 7.13 m and the largest diameter on the geodesic dome is 6.35 m. The double doors leading into the observatory are 2.14 m tall and 1.84 m wide to easily move equipment. (b) A top view of the first floor shows a potential layout. There is a pair of double doors, one inner door and one outer door is designed to minimize the risk of light leakage to the rest of the observatories. On the bottom right, a spiral stairwell with a door on the first floor is designed to minimize the light to the telescopes. Under the stairwell a small utility closet may be used for storage. The main work is light tight to allow engineering during the night.

## 5. SUMMARY

The PANOSETI observatory will observe 4,441 sq. deg. instantaneously at nano-to-second timescales. We have discussed the overall optical and mechanical conceptual design of the telescope and experimental apparatus. Each telescope module is comprised of three main assemblies: the 0.48 m Fresnel lens assembly, telescope housing, and electronics package that includes the detector plane. The Fresnel lens assembly consists of a retainer ring, protective cover, Fresnel lens, two support rods, and a Fresnel lens frame. The Fresnel lens assembly is attached to the telescope module housing. The telescope housing consists of an upper conical housing, top baffle, mounting flange, lower housing, and three spring loaded clips. The electronics assembly is installed at the bottom of the conical housing and held in place by the three spring loaded clips. The electronics assembly consists of a shutter assembly, focus stage, and the detector and electronics assembly.

There are forty-five modules per observatory, and they will each be attached to a triangular section of a tessellation 3 geodesic dome. Each module will be attached at three points with a lead screw system to allow for minor directional adjustment to accommodate manufacturing tolerances and alignment to the secondary sight. The geodesic dome is enclosed in a clam shell enclosure. The geodesic dome is attached to a concrete circular structure that supports the observatory. The main building is designed to have a first floor that will be used as a work space and house the electronics rack equipment that operates the observatory.

ACKNOWLEDGMENTS

The PANOSETI research and instrumentation program is made possible by the enthusiastic support and interest by Franklin Antonio. We thank the Bloomfield Family Foundation for supporting SETI research at UC San Diego in the CASS Optical and Infrared Laboratory. Harvard SETI is supported by The Planetary Society and The Bosack/Kruger Charitable Foundation. UC Berkeley's SETI efforts involved with PANOSETI are supported by NSF grant 1407804, the Breakthrough Prize Foundation, and the Marilyn and Watson

Alberts SETI Chair fund. We thank the staff at Mt. Laguna and Lick Observatories for their help with equipment testing. Lastly, we thank David Brotherson at Astrohaven for the helpful discussions about clam-shell enclosures.